\documentclass[a4paper]{article}
\usepackage{INTERSPEECH2020,amsmath,graphicx}
\usepackage{makecell}
\usepackage{url}
\usepackage{adjustbox}
\usepackage{float}

\title{Detecting and analysing spontaneous oral cancer speech in the wild}
\name{Bence Mark Halpern$^{123}$, Rob van Son$^{13}$, Michiel van den Brekel$^{13}$, Odette Scharenborg$^2$}
\address{$^1$University of Amsterdam, ACLC, Amsterdam, The Netherlands \\
    $^2$Multimedia Computing Group, Delft University of Technology, Delft, The Netherlands \\
  $^3$Netherlands Cancer Institute, Amsterdam, The Netherlands}

\email{b.halpern@nki.nl, r.v.son@nki.nl,  M.W.M.vandenBrekel@uva.nl, o.e.scharenborg@tudelft.nl}

\usepackage{todonotes}
\begin{document}
%
\maketitle
\begin{abstract}
Oral cancer speech is a disease which impacts more than half a million people worldwide every year. Analysis of oral cancer speech has so far focused on read speech.
In this paper, we 1) present and 2) analyse a three-hour long spontaneous oral cancer speech dataset collected from YouTube. 3) We set baselines for an oral cancer speech detection task on this dataset. The analysis of these explainable machine learning baselines shows that sibilants and stop consonants are the most important indicators for spontaneous oral cancer speech detection.
\end{abstract}
\noindent\textbf{Index Terms}: pathological speech, corpus, oral cancer speech, explainable AI

\section{Introduction}
\label{sec:intro}

Oral cancer is a disease which impacts approximately 529,500 people worldwide every year \cite{shield2017global}. Apart from improving survival rates (\textit{mortality}), research attention has shifted to improving the quality of life after surgery \cite{Epstein1999}. Oral cancer survivors can suffer from several problems affecting their quality of life: difficulty swallowing \cite{ward2014head,logemann1997speech}, decreased tongue mobility \cite{kappert2019quantification} and impaired speech intelligibility \cite{ward2014head}. The latter is the focus of our paper.

Speech impairment occurs due to the oral cancer treatment in which parts of the tongue or the entire tongue is removed (\textit{partial/total glossectomy}). This partial or full removal causes an inability to reach articulatory targets. Oral cancer speech is consequently primarily impaired at the articulatory level, while only patients who have also undergone radiation therapy also have problems with phonation \cite{ward2014head}.

Different studies show different characteristics of oral cancer speech impairment. Stop consonants (mainly /k/, /g/, /b/, /p/, /t/, /d/) \cite{bressmann2009speech,bressmann2004consonant} and alveolar sibilants (i.e., /s/, /z/) \cite{laaksonen2011longitudinal} seem to be primarily affected. In certain cases, patients are able to learn articulatory compensation techniques to adjust for the lost tongue tissue.
For example, /t/ and /d/ can be produced by an altered bilabial seal \cite{ward2014head}. The effect of glossectomy on vowels and diphthongs is less clear \cite{laaksonen2010speech, de2009objective}.

Analysis of oral cancer speech has so far focused on read speech. In the studies above, participants were asked to read a text passage. However, it has been shown that such structured tasks can fail to identify some characteristics of speech \cite{thomas2005automatic}. 

So far, no research has been carried out investigating whether oral cancer speech can be reliably differentiated from non-oral cancer speech automatically. The aim of the paper is three-fold: 1) we investigate whether spontaneous oral cancer speech can be differentiated from healthy speech, focusing on spontaneous speech for the first time as far as we know, and as such present the first baselines for oral cancer speech detection. 2) In order to do so, we collected a large dataset, which allows us to use machine learning techniques. Creating a large dataset of pathological speech is time-consuming due to slow patient recruitment. We, therefore, created a database of ``found'' oral cancer speech, which is freely available to the community.\footnote{\url{http://doi.org/10.5281/zenodo.3732322}} 3) We provide a preliminary analysis of the differences in spontaneous oral cancer speech and healthy speech.

Pathological speech detection is a broad field. There are two main approaches employed in the field. The first one is to develop a new acoustic feature using some knowledge about the pathological speech and use that in a simple classification model. Effectively, this is solving the problem in a divide-and-conquer approach: detecting known characteristics of a pathology and then feeding it into a classifier. A typical example of this is looking at unsuccessful phone realisations with an automatic speech recogniser (ASR) \cite{noth_2009, maier2006fully}. The second approach is to generate some acoustic features from the audio using standard feature extractors (frontends) like openSMILE \cite{eyben2015opensmile}, Kaldi \cite{povey2011kaldi} or librosa \cite{mcfee2015librosa}. This is a good approach if we are unsure what features would be the most appropriate. These features are then used to train a few chosen machine learning models (backends) such as artificial neural networks \cite{klumpp2018ann}, Gaussian mixture models \cite{dibazar2002feature}, support vector machines \cite{bocklet2008age} and boosted regression trees \cite{valentini2012automatic}. These techniques rely on the models' learning capabilities to find
any difference in the feature distributions of healthy and pathological speakers. We follow the second approach here, by using the Kaldi feature extractor along with ASR features.

In order to analyse the differences between oral cancer speech and healthy speech we only use backends which have some degree of explainability. Moreover, in addition to Kaldi features, we use phonetic posteriorgrams (PPG) as ASR-based features, which are easier to interpret than MFCC or PLP.


\section{Dataset}
\label{sec:dataset}

We manually collected audio data containing English, spontaneous oral cancer speech from 3 male speakers and 8 female speakers from YouTube. The presence of oral cancer speech in the audio was determined by the content of the video and the authors' experience with such speakers. The audio was manually cut to exclude music, healthy speakers and artefacts, leaving only the oral cancer speech. The total duration of the oral cancer dataset is 2h and 59mins. 

As our spontaneous healthy speech, we chose a subset of native American English speakers from the VoxCeleb dataset \cite{nagrani2017voxceleb}. This dataset was chosen because it was also originally collected from YouTube. This allows exclusion of YouTube characteristics as a confounding factor in the detection task. The gender and number of speakers, as well as the amount of speech material for each speaker, was matched with that of the speech of the 11 oral cancer speakers to ensure that the ratio of the recordings is similar in the two datasets. There is no overlap in speakers between the training and test sets. In total, there are 10 speakers (8 female, 2 male) in the training set, and there are 12 (8 female, 4 male) speakers in the test set. The total duration of the training set is 4 h and 36 mins, for the test set 1 h 28 mins. The average duration per speaker is 27.6 min in the training set and 7.3 min in the test set.

The recordings in the oral cancer dataset were automatically cut into 5 s chunks to match the average duration of the utterances in the VoxCeleb dataset using \texttt{ffmpeg}. The audio was downsampled to 16 kHz and converted from stereo to mono. Loudness was normalised to 0.1 dB using the \texttt{sox} tool.

%

\section{Method}
\label{sec:method}

We compared several frontend and backend combinations to find the best oral cancer vs. healthy speech detection system. Sections \ref{section:frontend} and \ref{section:backend} describe the different frontends and backends, respectively, and the rationale why we chose them. The code of the analysis is also available online\footnote{\url{https://github.com/karkirowle/oral_cancer_analysis}}.


\subsection{The preprocessing frontends}
\label{section:frontend}

The following features have been extracted from the audio (abbreviations in bold). All features were calculated using the Kaldi frontend \cite{povey2011kaldi}, unless mentioned otherwise.
Silences were cut using Kaldi's voice activity detection (VAD) algorithm. 

\begin{itemize}
\item \textbf{MFCC} - Mel-frequency cepstral coefficients are used as the baseline feature.
    \item \textbf{LTAS} - Long term average spectrum is used as a voice quality measurement in the early detection of pathological speech \cite{Smith2014,Master2006} and to
evaluate the effect of speech therapy or surgery on voice quality \cite{Tanner2005}. The LTAS features are extracted by calculating the mean and standard deviation of the frequency bins of Kaldi spectrograms and stacking them together.
    \item \textbf{PLP} - Perceptual linear predictive coefficients are known to be related to the geometry of the vocal tract based on the principles of source-filter theory \cite{fant1981source}. During oral cancer the geometry of the vocal tract changes, so we expect that PLPs have useful information for detection. The PLPs are calculated based on \cite{hermansky1990perceptual}.
    \item \textbf{Pitch} - To investigate whether there are also prosodic and phonation impairments in oral cancer speech, a combination of pitch and voicing likelihood feature is used \cite{ghahremani2014pitch}. 
    \item \textbf{PPG} - Phonetic posteriorgrams were calculated using an ASR trained on Librispeech \cite{panayotov2015librispeech}, based on the implementation of \cite{zhao2019foreign}. PPGs are probability distributions over a set of phones, i.e., what is the probability that this phone is spoken at this frame of the utterance. The implementation that we used included 40 phones, including the phone for silence. However, silence phones were excluded in our approach.
\end{itemize}

\subsection{The backends}
\label{section:backend}

Two different backends were used: a Gaussian Mixture Model (GMM) and a linear regression method (LASSO) \cite{tibshirani1996regression}. Linear regression is generally the easiest to interpret, however when the dimensionality of the features are high, a feature selection step is usually recommended, that is why we used LASSO. The GMM is used widely in pathological speech detection \cite{valentini2012automatic,dibazar2002feature}. The GMM and LASSO models were implemented using the \texttt{sklearn} \cite{sklearn_api} library.

In addition to these two traditional models, we also trained a Dilated Residual Network (ResNet) \cite{DBLP:journals/corr/YuKF17}. Similar architectures have been successful in detecting spoofed speech \cite{halpern_2020,Lai2019a}. We expect ResNet's ability to recognise unnatural speech to be useful for detection of pathological speech.

\subsubsection{Gaussian mixture model}

We trained separate GMMs for oral cancer speech and healthy speech. The number of mixture components for each GMM was chosen so that it maximises performance on the test set from the list of \( m = [4, 8, 10, 12, 16] \). This could result in overfitting to the test set, however, in practice we found that the test set performance is relatively insensitive to the choice of the mixture parameter. This is further discussed in Section \ref{sec:discussion}. We report the number of mixture components used with the results in Table \ref{tab:results}. At test time, we presented the healthy and the oral cancer speech utterances to both models. To determine whether the input speech frame contained healthy or oral cancer speech, we calculated the likelihoods for each speech frame and averaged over all frames to compute a single likelihood for the entire stretch of speech. The average likelihoods for both models were subsequently compared.

\subsubsection{LASSO}

LASSO is a variant of linear regression, which performs feature selection and regression simultaneously. It might be the case that for a given linear regression task, some features do not contain any relevant information to make predictions. In LASSO, coefficients of regression are encouraged to be close to zero if they do not provide useful information. Zeroing (pruning) some features means that the model requires only a subset of all predictors, making it parsimonious and easier to interpret. Pruning of the features is facilitated by setting the hyperparameter \( \alpha \): the larger this parameter is, the closer the coefficients are to zero. This hyperparameter is taken from the list \( \alpha = [0.1, 0.01, 0.001, 0.0001] \) and tuned on the test set (see Section \ref{sec:discussion}). The hyperparameters are reported with the results in Table \ref{tab:results}.

\begin{table*}[t]
  \caption{Equal error rates (EER) and accuracy of the classifiers with different feature and classifier combinations. Higher accuracy and lower EER is better. Best performances are emphasised in \textbf{bold}.}
  \vspace*{-3mm}
\begin{center}
\begin{adjustbox}{totalheight=100pt}
\newcolumntype{?}{!{\vrule width 2pt}}
\begin{tabular}{ | c | c | c | c | c | c ? c |}
\hline
 \textbf{GMM} & \textbf{PLP} & \textbf{LTAS} & \textbf{PPG} & \textbf{Pitch} & \textbf{MFCC} & \textbf{Spectrogram-ResNet} \\
 \hline
   Train set accuracy & 97.80\% & 94.71\% & 85.24\% & 52.04\% & 97.02\% & 98.58\% \\
    \hline
  Train set EER & 1.34\% & 5.3\% & 11.56\% & 39.07\% & 2.05\% & 1.00\% \\
    \hline
 Test set accuracy & 77.52\% & 65.59\% & 72.89\% & 43.57\% & 76.83\% & \textbf{88.37}\% \\ 
 \hline
 Test set EER & 22.01\% & 31.05\% & 29.33\% & 45.65\% & 20.62\% & \textbf{9.85}\% \\
 \hline
 \( m \) & 8 & 10 & 10 & 16 & 8 & N/A \\
 \hline
 \Xhline{5\arrayrulewidth}
  \textbf{LASSO} & \textbf{PLP} & \textbf{LTAS} & \textbf{PPG} & \textbf{Pitch} & \textbf{MFCC} & - \\ 
  \hline
  Train set accuracy & 85.46\% & 98.55\% & 80.59\% &  70.22\% & 87.02\% & - \\
    \hline
  Train set EER & 9.25\% & 1.45\% & 12.25\% & 29.03\% & 8.01\% & - \\
    \hline
  Test set accuracy & 80.19\% & 87.37\% & 73.35\% & 58.86\% & 80.88\% & - \\
    \hline
  Test set EER & 20.62\% & 10.67\% & 25.84\% & 37.32\% & 19.23\% & - \\
 \hline
  \( \alpha \) & 0.0001 & 0.01 & 0.001 & 0.001 & 0.01 & - \\
 \hline
\end{tabular}
\end{adjustbox}
\end{center}
 \label{tab:results}
\end{table*}

\subsubsection{Neural network classifier (Spectrogram-ResNet)}
\label{sec:architecture}
The ResNet architecture consists of four Dilated ResNet blocks. Each block has a different kernel size (width \(\times\) height) and number of filters: $(240\times100)$ and $8$; $(120\times 200)$ and $16$; $(60\times100)$ and $32$; $(30\times50)$ and $64$. This is followed by a fully connected layer with 100 hidden nodes and finally a softmax layer. The architecture is described in detail in in \cite{halpern_2020}. 


The input of the ResNet consisted of spectrograms. Spectrograms are highly informative, high dimensional features, which capture most properties of the raw speech signals. They are widely used with neural network backends \cite{shen2018natural,  Lai2019a}. The input spectrograms were zero padded to the length of the longest utterance so that even the longest utterance could be processed by the network. The network was trained for 50 epochs in Keras \cite{Chollet2015a}, selecting the model with the best validation loss after training. We used the Adam optimiser with a learning rate of \( \mu = 0.001 \) \cite{kingma2014adam}. To avoid overfitting on the test material, no hyperparameter optimisation was performed (see Section \ref{sec:discussion}).






\section{Results and analysis}
\label{sec:results}


\subsection{Results on training and test set}

The detection accuracies for the training and test sets are measured using accuracy and equal error rate (EER), and can be seen in Table \ref{tab:results}. Chance level accuracy for the test set is 57.82\%. \( m \) refers to the number of Gaussian mixture components used during GMM training. \( \alpha \) refers to the sparsity inducing hyperparameter of LASSO, a larger \( \alpha \) indicates a sparser model.

The Spectrogram-ResNet-based detector achieved the best classification performance both in terms of accuracy and EER. This is closely followed by the LTAS-LASSO model. The superiority of the ResNet over the other methods is  likely due to  the ResNet seeing the whole utterance at once unlike the other methods. LASSO backends always outperformed the GMM-based backends on the test set, which is especially striking on the LTAS-based features. One possible explanation for the performance difference might be the collinearity of certain features as LASSO is known to handle collinearity better. We can see that for non-collinear features like MFCC or PLP the performance difference between LASSO and the GMMs is marginal.
The worst performance was achieved by the Pitch features, which is close to chance level for both backends. This indicates that Pitch features are not appropriate for oral cancer speech detection, suggesting that oral cancer speech indeed is impaired primarily on the articulation level \cite{ward2014head}.


\subsection{Analysis of the differences between oral cancer speech and healthy speech}
\label{section:analysis}


To investigate the differences between oral cancer speech and healthy speech the two best performing architectures (Spectrogram-ResNET, LTAS-LASSO) and PPG-GMM were analysed through the information in the speech signal that was used by these models to distinguish oral cancer speech from healthy speech.
While the PPG-GMM does not stand out in terms of accuracy, it lends itself to easy interpretation of the acoustic information used for the task.

\subsubsection{Analysis of the Spectrogram-ResNet detector}

To investigate what information the ResNet classifier uses to distinguish between oral cancer speech and healthy speech, we look at what parts of the spectrogram change the classification results the most. 

To find these salient parts of the spectrogram, we calculate mean class activation maps, which indicate what frequencies are the most important for detection of both classes, for each sample in our test set as follows: Given a spectrogram image and a class label (oral cancer speech/healthy speech) as input, we pass the image through the ResNet to obtain the raw class scores before softmax. The gradients are set to zero for all classes except the desired class (i.e., oral cancer speech), which is set to 1. This signal is then backpropagated to the rectified convolutional feature map of interest \cite{Selvaraju2017a}. We used the implementation from the 
\texttt{keras-vis} library.

\begin{figure}
\begin{center}
\includegraphics[width=8cm]{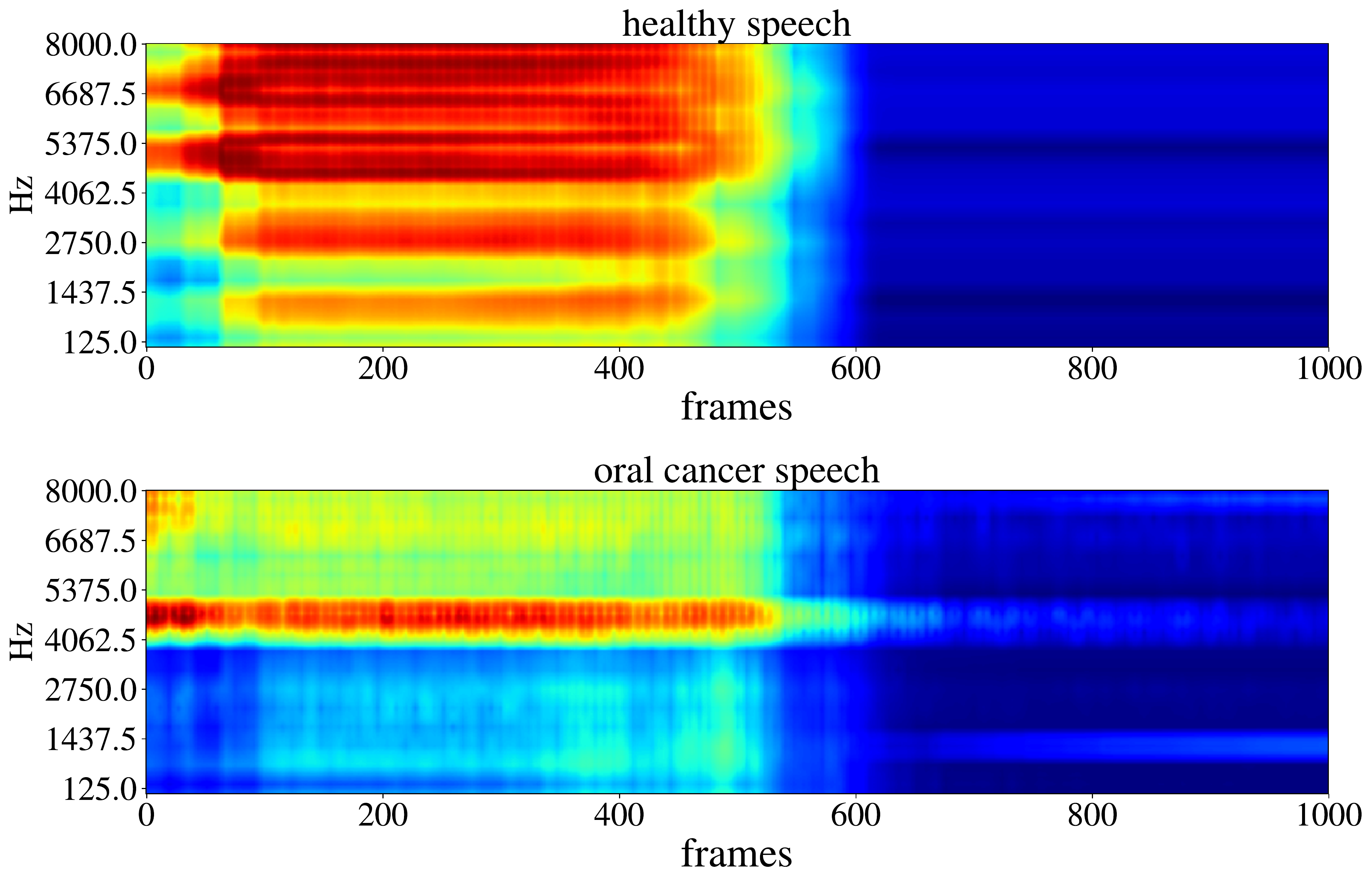}
\end{center}
  \vspace*{-3mm}
\caption{Mean class activation maps for healthy speech (top panel) and oral cancer speech (bottom panel).}
\label{fig:act}
\end{figure}

Figure \ref{fig:act} shows the mean class activation maps for healthy speech and oral cancer speech. For healthy speech (top panel), the neural network spreads its focus (indicated by the coloured bands where red means higher intensity) among all the frequency bands. For oral cancer speech (bottom panel), the majority of the acoustic energy lies above the 4 kHz band. This indicates that sibilant frequencies, which are above 4 kHz \cite{Ladefoged2012}, might be important for distinguishing between oral cancer and healthy speech. This is in agreement with previous studies who name sibilants as impaired sounds  \cite{laaksonen2011longitudinal}.

\subsubsection{Analysis of the PPG-GMM detector}

The trained GMM models can be viewed as models of a global oral cancer speaker and a global healthy speaker. The mean parameters of the GMMs can inform us what features are more typical for oral cancer speakers and which for healthy speakers by constructing a difference model. First, we calculate the difference of the mixture components in the two GMMs, obtaining a \( d \in \mathbb{R}^{m} \) difference vector for each phone. Taking the mean of \( d \), we are able to obtain a signed scalar \( p \in \mathbb{R} \) for each phone class. If \( p \) is positive it means that there is a higher likelihood of occurrence of that phone in oral cancer speech compared to healthy speech. If \( p \) is negative it means that the likelihood of that phone is lower in oral cancer speakers -- meaning that they have trouble pronouncing that phone.

\begin{figure}
    \centering
    \includegraphics{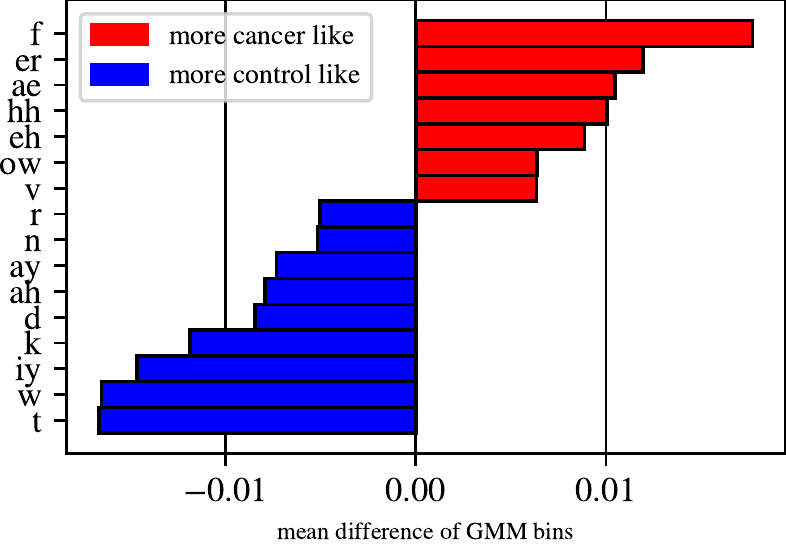}

    \caption{Mean difference of GMM bins \( (p) \) of the PPG-GMM architecture.}
    \label{fig:gmm_ppg_barplot}

\end{figure}

Figure \ref{fig:gmm_ppg_barplot} shows the results for the phones with absolute mean differences larger than 0.005. The blue bars indicate phones that are more often present in healthy speech and the red bars indicate phones that are more typical of cancer speakers. We can see that /t/, /w/, /iy/, /k/ and /d/ have lower likelihoods, indicating that some stop consonants are challenging for oral cancer speakers. This is in agreement with \cite{bressmann2009speech,bressmann2004consonant}.

\subsubsection{Analysis of the LTAS-LASSO based detector}

\begin{figure}
    \centering
    \includegraphics{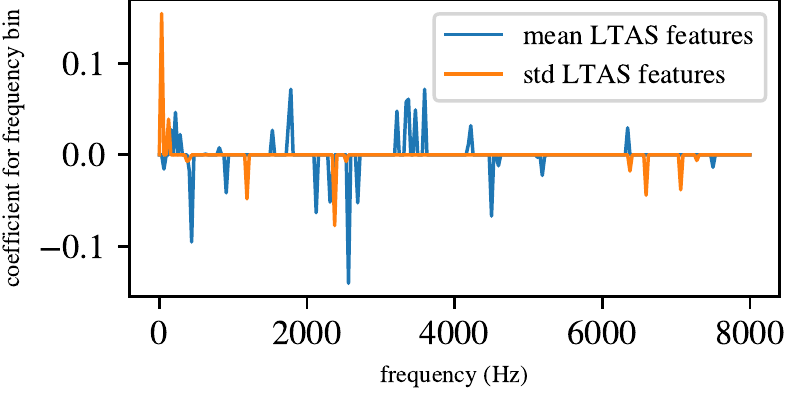}
    \caption{Learned coefficients of the LTAS-LASSO.}
    \label{fig:lasso_ltas_barplot}
\end{figure}

LASSO-based models can be analysed using the coefficients of regression. 
A positive coefficient indicates a feature contributing to the cancer class and vice versa. Figure \ref{fig:lasso_ltas_barplot} shows the learned coefficients. The blue line shows the mean coefficients, the orange line shows the standard deviation coefficients of the LTAS. Knowing that neighbouring frequencies are discouraged (because they provide similar (collinear) information), it is still surprising that some frequency bands have several adjacent positive/negative coefficients (clusters, shown as adjacent spikes). These clusters indicate that a greater level of frequency resolution is needed for that particular frequency band. We can see that for oral cancer speech this is the 3-4 kHz, indicating that sibilant frequencies need greater frequency resolution. 



\section{General discussion}
\label{sec:discussion}

The paper presents the first baseline models for the task of healthy vs. oral cancer speech classification. The Spectrogram-ResNet classifier achieved a high classification accuracy and outperformed all other models.
A preliminary analysis of the three models indicated that sibilant frequencies, and stop consonant phones are the most decisive in the classification.

The current study used healthy speech from the VoxCeleb dataset, which only contains recordings from celebrities. Although potentially the detectors could use other features than those related to the acoustic characteristics of the speech for classification, this is not likely: inspection of the recognition results of the individual speakers in both datasets showed that in both datasets some speakers are well classified whereas others are not (range oral cancer speech: 49.7\% -- 100.0\%; range healthy speech: 34.8\% -- 94.4\% on the test set), although on average the oral cancer speakers were better classified than the healthy speakers:  89.6\% vs. 66.2\%.



Because of our relatively small-sized datasets, we kept hyperparameter tuning to a minimum to avoid overfitting on the test set. The more traditional methods (GMM and LASSO) only have a single hyperparameter, so chances of overfitting to the test set are small. Neural networks, on the other hand, usually have a myriad of hyperparameters, which makes overfitting more likely. To avoid this, we refrained from using any tuning mechanisms at all with the neural networks.


\section{Conclusion}
\label{sec:conclusion}

We presented a brand new dataset for analysis of spontaneous oral cancer speech, and showed that a detector based on ResNet taking spectrograms as input had a high performance in distinguishing between oral cancer speech and healthy speech, and  generalised well to unseen data. Analysis of the speech signals through the classifiers shows that sibilants and stop consonants are important for oral cancer speech detection, while no evidence has been found on the importance of vowels and diphthongs.


\section{Acknowledgements}
\label{sec:acknowledgements}

This project has received funding from the European Union’s
Horizon 2020 research and innovation programme under Marie
Sklodowska-Curie grant agreement No 766287. The Department of Head and Neck
Oncology and surgery of the Netherlands Cancer Institute
receives a research grant from Atos Medical (H\"orby, Sweden),
which contributes to the existing infrastructure for quality of
life research.

\bibliographystyle{IEEEbib}
\bibliography{main}

\end{document}